\documentclass[final,5p,times]{elsarticle}
\usepackage{textcomp}
\usepackage{graphicx}
\usepackage{amsmath}
\usepackage{amssymb}
\usepackage[usenames]{color}
\usepackage{url}
\biboptions{square,comma,sort&compress}
\usepackage{hyperref}
\journal{Journal of Crystal Growth}

\begin{document}
\begin{frontmatter}

\title{Structural Investigation of InAs-AlInAs and InAs-AlInP Core-Shell Nanowires}
\author[mac]{C. M. Haapamaki}
\author[iqc,watchem]{J. Baugh}
\author[mac]{R. R. LaPierre\corref{lapierre}}
\ead{lapierr@mcmaster.ca}
\address[mac]{Department of Engineering Physics, Centre for Emerging Device Technologies,
McMaster University, Hamilton, Ontario, L8S 4L7, Canada}
\address[iqc]{Institute for Quantum Computing, University of Waterloo, Waterloo, Ontario, N2L 3G1, Canada}
\address[watchem]{Department of Chemistry, University of Waterloo, Waterloo, Ontario, N2L 3G1, Canada}
\cortext[lapierre]{Corresponding author}

\begin{abstract}
InAs nanowires were grown on GaAs substrates by the Au-assisted vapour-liquid-solid (VLS) method in a gas source molecular beam epitaxy (GS-MBE) system.  Passivation of the InAs nanowires using InP shells proved difficult due to the tendency for the formation of axial rather than core-shell structures.   To circumvent this issue, Al$_x$In$_{1-x}$As or Al$_x$In$_{1-x}$P shells with nominal Al composition fraction of x = 0.20, 0.36, or 0.53 were grown by direct vapour-solid deposition on the sidewalls of the InAs nanowires.  Characterization by transmission electron microscopy revealed that the addition of Al in the shell resulted in a remarkable transition from the VLS to the vapour-solid growth mode with uniform shell thickness along the nanowire length.  Possible mechanisms for this transition include reduced adatom diffusion, a phase change of the Au seed particle and surfactant effects.  The InAs-AlInP core-shell nanowires exhibited misfit dislocations, while the InAs-AlInAs nanowires with lower strain appeared to be free of defects.  
\end{abstract}


\end{frontmatter}

\section{Introduction}\label{sec:intro}
Semiconductor nanowires have commanded tremendous attention in the nanoscience and nanotechnology community in the past decade.  InAs nanowires, in particular, are of interest for the continued down-scaling of MOSFET devices due to the high electron mobility of InAs and the possible integration of these nanowires with Si.  In addition, the small bandgap of InAs makes this material of interest in infrared photodetectors ~\cite{Wei2009}.  However, optimal device performance requires the removal of surface states (i.e., passivation) to prevent ionized impurity or surface roughness scattering of electrons in InAs nanowire MOSFET devices, and to prevent electron-hole carrier recombination and depletion in photodetector or photovoltaic devices.  Passivation of InAs can be achieved using a thin (several nm) shell of InP surrounding the InAs core.  Core-shell InAs-InP nanowire heterostructures have demonstrated improved performance in MOSFET devices compared to their unpassivated counterparts \cite{Dayeh2007,Jiang2007,Ford2009,Tilburg2010}.  InAs-InP core-shell structures also provide quantum confinement of carriers \cite{Zanolli2007}, which is useful for efficient radiative recombination in single photon sources, for example \cite{Borgstrom2005}.
  
InAs-InP core-shell heterostructures can be achieved using metalorganic vapour phase epitaxy (MOVPE).  In MOVPE, InP shell growth can be promoted compared to axial growth by increasing the growth temperature. At elevated temperatures pyrolysis of the metalorganic precursors occurs on the nanowire sidewalls resulting in lateral vapour-solid deposition \cite{Mohan2006,Masumoto2011,Masumoto2010,Li2007,Tchernycheva2007}.  An alternative technique for nanowire growth is molecular beam epitaxy (MBE) where growth is dominated by diffusive transport of adatoms from the substrate surface or nanowire sidewalls.  Thus, radial (shell) growth of nanowires in MBE can occur by kinetically limiting adatom diffusion resulting in vapour-solid deposition on the nanowire sidewalls.  However, InAs-InP core-shell structures are difficult to achieve in MBE because the diffusion length of indium is typically greater than the nanowire length (on the order of several microns); hence, axial rather than core-shell structures are typically achieved in MBE-grown InAs-InP nanowires.  To overcome the difficulty of achieving passivation of InAs nanowires by MBE, an alternative passivation scheme is examined consisting of AlInAs or AlInP shells.  In this paper, we show that shell growth can be significantly increased in MBE simply by the addition of Al in the shell composition.

\section{Experimental Details}\label{sec:exp}
Zn-doped GaAs (111)B (1-5$\times$10$^{18}$ cm$^{-3}$) wafers were treated by UV-ozone oxidation for 20 min to remove any hydrocarbon contamination and to grow a sacrificial oxide. The wafers were then dipped in buffered HF for 30 s and rinsed under flowing deionized water for 10 min.  The wafers were dried with N$_2$ and transferred to an electron beam evaporator where a 1 nm thick Au film was deposited at room temperature at a rate of 0.1 nm/s as measured by a quartz crystal thickness monitor.  The wafers were then transferred to a gas source molecular beam epitaxy (GS-MBE) system (SVT Associates). In GS-MBE, group III species (In, Al) are supplied as monomers from an effusion cell, and group V species (P, As) are supplied as P$_2$ and As$_2$ dimers that are cracked from PH$_3$ and AsH$_3$ in a gas cracker operating at 950 $^{\circ}$C. Prior to growth, the wafers were placed in a pre-deposition chamber where they were degassed for 15 min at 300 $^{\circ}$C.  After transferring the wafers to the growth chamber, an oxide desorption step was performed where they were heated to 540 $^{\circ}$C under inductively coupled hydrogen plasma and As$_2$ overpressure for 10 min, leading to the formation of Au nanoparticles. The sample was cooled to the growth temperature of 420 $^{\circ}$C and nanowire growth was initiated by opening the In shutter.  InAs was grown for 15 min at an equivalent 2D growth rate of 0.14 nm/s. After growth of the InAs, the growth was switched to InP, Al$_x$In$_{1-x}$As or Al$_x$In$_{1-x}$P with different Al alloy fractions x, calibrated by prior 2-D depositions (see Table~\ref{tab:growths}).  The shells were grown for a duration of 7.5 min at a 2-D equivalent rate of 0.14 nm/s.

\begin{table}[!ht]
\begin{center}
 \begin{tabular}{ | l || c |}
 \hline
Segments Grown 	& Lattice Mismatch Strain\\
&  (\%)\\
\hline
InAs – InP	& 3.1 \\
InAs - Al$_{0.20}$In$_{0.80}$As &	1.3 \\
InAs - Al$_{0.36}$In$_{0.64}$As &	2.4 \\
InAs - Al$_{0.53}$In$_{0.47}$As &	3.5 \\
InAs - Al$_{0.36}$In$_{0.64}$P &	5.5 \\
InAs - Al$_{0.53}$In$_{0.47}$P &	6.6 \\
\hline
 \end{tabular}
\caption{\label{tab:growths}Description of samples}
\end{center}
\end{table}  

The nanowire morphology was determined using a JEOL 7000F field emission scanning electron microscope (SEM). The nanowires were removed from the growth substrate by sonication in methanol solution for structural investigation.  The solution was then drop cast onto a holey carbon grid for examination using a JEOL 2010F high resolution scanning transmission electron microscope (HRTEM) using high angle annular dark field (HAADF) and selected area diffraction (SAD) modes.  The chemical composition was determined by energy dispersive x-ray spectroscopy (EDX).

\section{\label{sec:results}Results and Discussion}
\begin{figure}[!ht]
\begin{center}
 \includegraphics[width=\columnwidth]{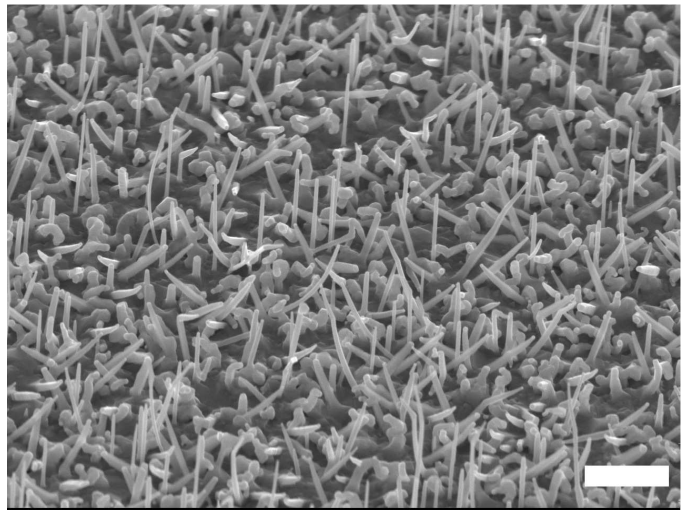}
  \caption{\label{fig:fig1}45$^{\circ}$ tilted SEM image of InAs-Al$_{0.20}$In$_{0.80}$As sample.  Scale bar is 1 $\mu$m.}
\end{center}
\end{figure}

A representative SEM image of the nanowires is shown in Figure~\ref{fig:fig1} for the InAs-Al$_{0.20}$In$_{0.80}$As sample.  Similar results were obtained for the other samples in Table~\ref{tab:growths}.  Close inspection of the tip of the nanowires always indicated the presence of a Au particle consistent with the VLS process.  Based on the SEM analysis, the nanowires were roughly 1 $\mu$m in length and 40 nm in diameter on average with a rod-shaped morphology.  Many of the nanowires grew orthogonal to the substrate surface, although tilted growth directions were also observed which may have occurred due to the lattice mismatch strain of the InAs nanowires relative to the GaAs substrate \cite{Fortuna2010}.  Further work is required to optimize the growth conditions to achieve a high density of vertical nanowires with monodisperse diameter.

\begin{figure*}[!ht]
\begin{center}
 \includegraphics[width=1.5\columnwidth]{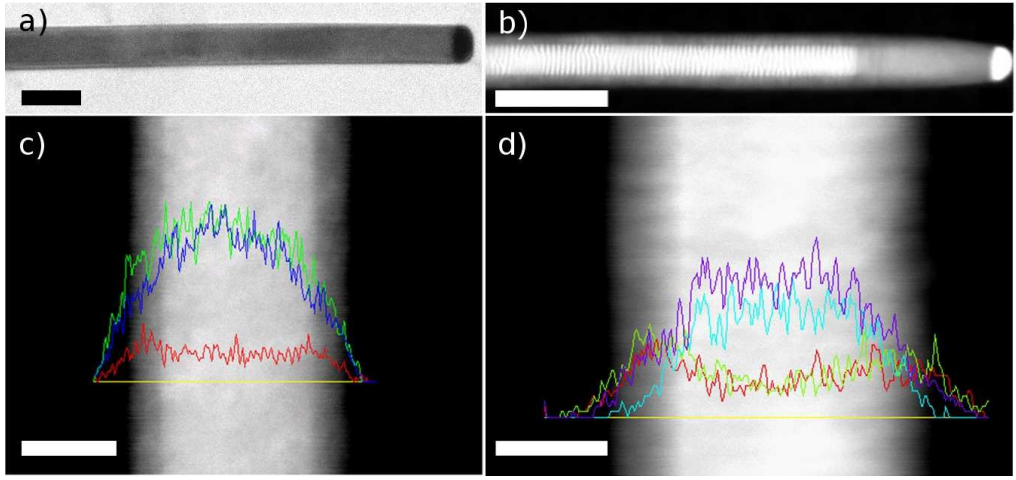}
  \caption{\label{fig:fig2}(Color online) (a) BF-TEM image of InAs nanowire without a shell. Scale bar is 100 nm. (b) HAADF image of InAs-Al$_{0.53}$In$_{0.47}$P core-shell nanowire.  Scale bar is 100 nm.  (c) HAADF image of InAs-Al$_{0.20}$In$_{0.80}$As core-shell nanowire with superimposed EDX linescan (Al: red, In: blue, As: green).  Scale bar is 20 nm. (d) HAADF image of InAs-Al$_{0.53}$In$_{0.47}$P core-shell nanowire with superimposed EDX linescan (Al: red, In: purple, As: turquoise, P: green).  Scale bar is 20 nm.}
\end{center}
\end{figure*}

Bare InAs nanowires, grown without any InP, AlInAs or AlInP segments, exhibited a constant diameter, negligible tapering, and a Au particle diameter roughly equal to the nanowire diameter as shown in the representative TEM image of Figure~\ref{fig:fig2}(a).  These observations indicate that InAs nanowires grew in the axial direction by the VLS process with negligible vapour-solid growth on the nanowire sidewalls.  Growth of an InP segment following the InAs segment also resulted in a purely axial heterostructure with negligible radial (shell) growth.  The detailed analysis of the axial InAs-InP heterostructures has been reported in detail elsewhere \cite{Haapamaki2011}.  

The purely axial growth of InAs or InAs-InP nanowires can be explained by the diffusion length of In adatoms exceeding the length of the nanowires.  In such a case, In adatoms can diffuse from the substrate or nanowire sidewalls to reach the Au particle at the top of the nanowire, contributing to axial VLS growth and negligible radial growth.  While such axial InAs-InP heterostructures are interesting in their own right, they do not permit the core-shell structure that is required for passivation.  To achieve a shell surrounding an InAs core, the alternative shell material consisting of AlInAs or AlInP was examined. 

The structure of the InAs-AlInAs and InAs-AlInP nanowires was examined by HAADF images where larger atomic mass appears as brighter contrast.  Figure~\ref{fig:fig2} shows representative HAADF images for the InAs-Al$_{0.20}$In$_{0.80}$As and InAs-Al$_{0.53}$In$_{0.47}$P nanowires.  Similar results were obtained for the other samples in Table~\ref{tab:growths}.  The lighter contrast near the nanowire axis indicates the InAs core, while the darker contrast along the nanowire edge indicates the presence of an AlInAs (Figure~\ref{fig:fig2}(c)) or AlInP (Figure~\ref{fig:fig2}(b, d)) shell.  This core-shell geometry was verified by EDS linescans across the nanowire diameter superimposed on the HAADF images.  Al content peaked at the edges of the nanowire in Figure~\ref{fig:fig2}(c) consistent with an AlInAs shell, while Al and P content peaked at the edges of the nanowire in Figure~\ref{fig:fig2}(d) consistent with an AlInP shell.  As observed in Figure~\ref{fig:fig2}(b), the shell thickness was remarkably uniform along the entire nanowire length.  

In contrast to bare InAs or InAs-InP nanowires which had a constant nanowire diameter equal to the Au particle diameter, the core-shell InAs-AlInAs and InAs-AlInP nanowires exhibited a slight tapering of the diameter near the top of the nanowire as observed in Figure~\ref{fig:fig2}(b).  This tapering indicates that the AlInAs and AlInP deposition occurred as vapour-solid deposition on the nanowire sidewalls over the InAs core, slightly increasing the nanowire diameter compared to the Au particle.  A certain degree of axial growth of the AlInAs or AlInP material was also observed vertically above the InAs core as seen by the darker contrast below the Au particle in Figure~\ref{fig:fig2}(b).  Measurement of these axial segments indicated lengths between 65 and 70 nm for all nanowires, which is near the expected contribution due to direct impingement on the Au particle (0.14 nm/s $\times$ 7.5 min = 63 nm).   Hence, during deposition of the AlInAs or AlInP, there appeared to be negligible contribution to the axial growth due to adatom diffusion from the nanowire sidewalls or substrate surface.  Comparing the latter results to the bare InAs or InAs-InP axial structures considered earlier, it is clear that the presence of Al in the shell results in a transition of the growth mode from axial to radial growth. 
The nominal shell thickness can be calculated as follows. Although the substrate is rotated during growth to ensure a uniform flux distribution on the surface, the flux sees a rectangle of height $L\sin\theta$ (equal to the nanowire height) and width $D$, where $\theta = 35^{\circ}$ is the angle of the molecular beam relative to the substrate surface, and $D$ is the nanowire core diameter. The number of atoms per unit time contributing to the axial growth is simply $JDL\sin\theta$ where $J$ is the impinging group III flux (nm$^{-2}$s$^{-1}$), which translates to a volume growth rate of $dV/dt=FDL\sin\theta$ where $F$ is the 2-D equivalent growth rate (nm s$^{-1}$).  The shell thickness $\delta R$ can be estimated by equating the latter volume to that of a shell wrapped around a cylindrical core of diameter $D$ \cite{Plante2009}:

\begin{equation}\label{eq:1}
tFDL\sin\theta=L \left [ \pi \left ( \frac{D}{2}+\delta R \right ) ^2 - \frac{\pi D^2}{4} \right ]
\end{equation}

where $t$ is the deposition time.  The left-hand expression of Eq.~\ref{eq:1} is the total volume of material collected by the nanowire sidewalls due to direct impingement while the right-hand expression is the volume of the resulting shell.  Rearranging Eq.~\ref{eq:1}, the thickness of the shell can be determined as:

\begin{equation}\label{eq:2}
\delta R = \frac{D}{2} \left ( \sqrt{\frac{4\delta R_o}{D}+1} -1 \right )
\end{equation}
				
where $\delta R_o = Ft\sin\theta /\pi$ is the shell thickness for large core diameters ($D >> \delta R_o$).  The independence of $\delta R_o$ with regard to nanowire length and diameter can be understood by the fact that the adatom collection area is proportional to the nanowire surface area ($L\pi D$), while the thickness of the shell is inversely proportional to surface area.  Using the shell growth conditions of $t$= 7.5 min, $F$=0.14 nm/s, and $\theta$= 35$^{\circ}$ gives $\delta R_o$= 11.5 nm.  Measurements by HRTEM in over 20 nanowires among the samples in Table~\ref{tab:growths} indicated an average core diameter and standard deviation of 33.6 $\pm$ 6.6 nm.  According to Eq.~\ref{eq:2}, this average core diameter corresponds to a shell thickness of 9.1 nm.   The average shell thickness and standard deviation measured directly by HRTEM, such as that shown in Figure~\ref{fig:fig2}, was 10.5 ± 1.3 nm, consistent with the above estimate.  

\begin{figure*}[!ht]
\begin{center}
 \includegraphics[width=1.5\columnwidth]{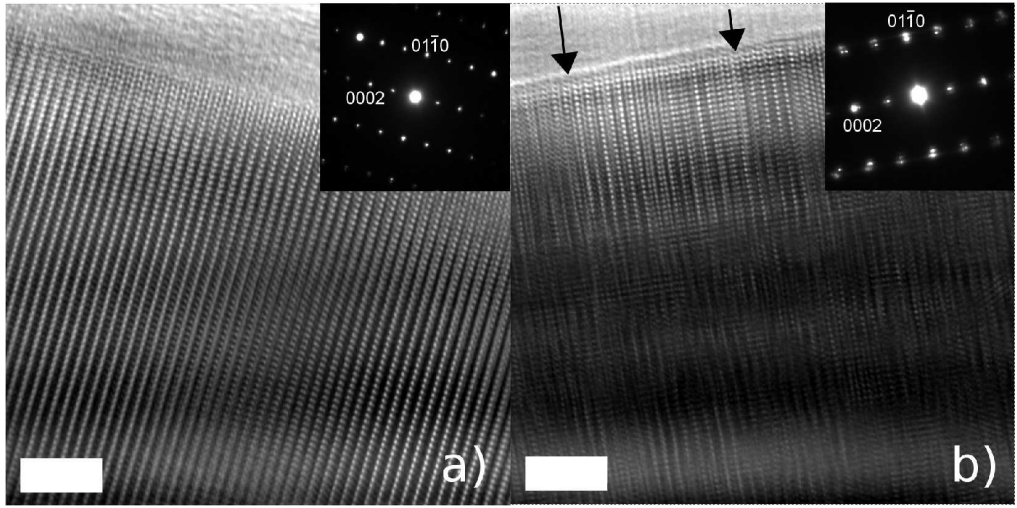}
  \caption{\label{fig:fig3}HRTEM along the (2$\bar{1}\bar{1}$0) zone axis for (a) InAs-Al$_{0.20}$In$_{0.80}$As and (b) InAs-Al$_{0.53}$In$_{0.47}$P core-shell nanowire.  Scale bars are 5 nm. Insets in (a) and (b) show SAD patterns.  Arrows in (b) indicate the positions of dislocations in the shell.}
\end{center}
\end{figure*}

As the Al content x in the shells increased, the lattice mismatch between the InAs core and the Al$_x$In$_{1-x}$As or Al$_x$In$_{1-x}$P shell increased (see Table~\ref{tab:growths}).  This increases the strain between the two materials and, if a critical thickness is exceeded, the shell will relax and misfit dislocations will be introduced \cite{Raychaudhuri2006,Trammell2008,Raychaudhuri2006JVSTB,Kastner2004,Gutkin2011,Fang2009}. This strain relaxation can be seen in the periodic Moir\'{e} fringes along the nanowire length such as that shown in Figure~\ref{fig:fig2}(b).  Assuming complete relaxation of the shell, the spacing between Moir\'{e} fringes can be predicted according to $L=a_1 a_2/(a_2-a_1)$ where $a_1$ and $a_2$ are the (111) lattice spacing of the shell and core, respectively.  For example, using $a_2$=6.0583/$\sqrt{3}$ {\AA} for the InAs core and $a_1$=5.6564/$\sqrt{3}$ {\AA} for the Al$_{0.53}$In$_{0.47}$P shell \cite{Vurgaftman2001} gives $L$= 4.9 nm, which is close to the measured value of $L$= 4.7 nm.  Similar Moir\'{e} fringes were observed in InAs-GaAs core-shell nanowires \cite{Paladugu2008,Paladugu2009,Kavanaugh2011}.  Moir\'{e} fringes were observed in all the AlInP samples indicating strain relaxation.   Moir\'{e} fringes were not evident in the AlInAs samples, which have lower strain compared to the AlInP samples (see Table~\ref{tab:growths}), suggesting only partial relaxation or pseudomorphic growth.  

Keplinger et. al. \cite{Keplinger2009} used grazing incidence x-ray diffraction to map a "phase diagram" of the transition from pseudomorphic to plastically relaxed shell growth for InAs-InAsP core-shell nanowires.  These results indicated that a pseudomorphic shell of 1.3\% strain (similar to Al$_{0.20}$In$_{0.80}$As) can have a thickness no greater than $\sim$20 nm, while a strain of 2.4\% (similar to Al$_{0.36}$In$_{0.64}$As) is expected to be plastically relaxed for all shell thicknesses.  Raychaudhuri et. al. \cite{Raychaudhuri2006,Raychaudhuri2006JVSTB} showed that an InAs nanowire core with InGaAs shell of 1.3\% strain (similar to Al$_{0.20}$In$_{0.80}$As) and 10 nm thickness will be dislocation-free for nanowire diameters less than 30 nm.  7-10 nm thick InP shells on InAs cores of 20 nm diameter (3.1\% strain) have shown no dislocations \cite{Li2007}.  Considering the above results, the Al$_{0.20}$In$_{0.80}$As shells are expected to be pseudomorphic, while the other AlInAs samples (Table~\ref{tab:growths}) may be pseudomorphic or partially relaxed.  The AlInP shells are expected to be completely relaxed.

HRTEM, such as the representative images in Figure~\ref{fig:fig3}(a, b), indicated that all nanowires exhibited the wurtzite crystal structure as commonly found in nanowires \cite{Dick2010}.  The wurtzite crystal structure of the nanowires was confirmed by SAD as represented in the inset of Figure~\ref{fig:fig3}(a) and (b).  A common occurrence in III-V nanowires is the existence of stacking faults whereby the crystal structure alternates between zincblende and wurtzite, or exhibits twinning, along the nanowire length \cite{Dick2010}.  Stacking faults in our core-shell nanowires occurred rarely (a few per nanowire) due to the low growth rate employed (0.14 nm/s), similar to the stacking-fault free GaAs nanowires described previously \cite{Plante2008,Plante2008Nt}.  Misfit dislocations were evident in the HRTEM images of the AlInP nanowires such as Figure~\ref{fig:fig3}(b) for the InAs-Al$_{0.47}$In$_{0.53}$P nanowire.  Spot splitting in the SAD pattern of InAs-AlInP nanowires, such as that shown in the inset of Figure~\ref{fig:fig3}(b), indicates strain relaxation, similar to previous reports for InAs-GaAs core-shell nanowires \cite{Kavanaugh2011}.  On the other hand, no misfit dislocations were evident in the HRTEM images of any AlInAs shells (for all samples in Table~\ref{tab:growths}), such as that in Figure~\ref{fig:fig3}(a) for an InAs-Al$_{0.20}$In$_{0.80}$As nanowire, consistent with the lack of spot splitting in the SAD patterns (e.g., inset of Figure~\ref{fig:fig3}(a)).

In the absence of Al, the adatom diffusion length of In evidently exceeds the nanowire length, resulting in purely axial growth of InAs and InP.  In contrast, AlInAs or AlInP deposition resulted predominantly in radial growth.  The addition of even a small amount of Al (x = 0.20) had a remarkable effect in reducing the VLS growth and, instead, promoted vapour-solid growth on the nanowire sidewalls.  Although the growth mechanisms responsible for this effect are not fully understood, the adatom diffusion appeared to be significantly reduced during shell growth.  Based on prior studies \cite{Fakhr2010,Chen2007}, the adatom diffusion length of In, Ga and Al are believed to consecutively decrease.  Al adatoms, in particular, are known to have a relatively short diffusion length, which has been used to create AlGaAs shells on GaAs cores \cite{Chen2007,Chen2008}.  In addition, the effect of Al on the phase of the Au seed particle is not currently known.  Based on the bulk phase diagrams \cite{OkamotoAuIn,OkamotoAlAu}, the growth temperature of 420 $^{\circ}$C is close to the Au-In eutectic temperature of 450 $^{\circ}$C but well below the Au-Al eutectic temperature of 525 $^{\circ}$C.  Hence, solidification of the Au particle and significant slowing of the axial growth due to changes in bulk solubility or diffusion may occur with the addition of Al.  However, it should be noted that any Al content in the Au particle was below the detection limit of EDS.  Finally, the addition of Al may have surfactant effects which reduces the In adatom surface diffusivity on the nanowire sidewalls \cite{Grandjean1996,Tournie1995}

\section{Conclusions}\label{sec:conc}
The present results are expected to extend MBE-grown nanowires into new device applications, for example InAs-AlInAs nanowires for MOSFET devices, similar to planar HEMTs \cite{Vasallo2007}.  Work is currently in progress to demonstrate the passivating properties of AlInAs for carrier transport in channels comprised of InAs nanowires.  Note that the InAs-AlInAs and InAs-AlInP material system are type I heterostructures suitable for carrier confinement \cite{Pistol2008}.  While the InAs-AlInP system exhibited misfit dislocations, the InAs-AlInAs system with low Al content appeared defect-free.  Our results are generally applicable to other nanowire systems such as the passivation of InP nanowires using lattice-matched Al$_{0.48}$In$_{0.52}$As, or the passivation of GaAs using lattice-matched Al$_{0.52}$In$_{0.48}$P.  

\section{Acknowledgements}\label{sec:ack}
Financial assistance from the Natural Sciences and Engineering Research Council of Canada and the Ontario Centres of Excellence is gratefully acknowledged.  Assistance with the TEM by Fred Pearson and with the MBE growths by Shahram Tavakoli are gratefully acknowledged.

\bibliographystyle{elsarticle-num} 
\bibliography{haapamaki_coreshell}

\begin{thebibliography}{10}
\expandafter\ifx\csname url\endcsname\relax
  \def\url#1{\texttt{#1}}\fi
\expandafter\ifx\csname urlprefix\endcsname\relax\def\urlprefix{URL }\fi
\expandafter\ifx\csname href\endcsname\relax
  \def\href#1#2{#2} \def\path#1{#1}\fi

\bibitem{Wei2009}
W.~Wei, X.-Y. Bao, C.~Soci, Y.~Ding, Z.-L. Wang, D.~Wang, {Direct Heteroepitaxy
  of Vertical InAs Nanowires on Si Substrates for Broad Band Photovoltaics and
  Photodetection}, Nano Letters 9~(8) (2009) 2926--2934.
\newblock \href {http://dx.doi.org/10.1021/nl901270n}
  {\path{doi:10.1021/nl901270n}}.

\bibitem{Dayeh2007}
S.~A. Dayeh, C.~Soci, P.~K.~L. Yu, E.~T. Yu, D.~Wang, {Transport properties of
  InAs nanowire field effect transistors: The effects of surface states},
  Journal of Vacuum Science and Technology B 25~(4) (2007) 1432--1436.
\newblock \href {http://dx.doi.org/10.1116/1.2748410}
  {\path{doi:10.1116/1.2748410}}.

\bibitem{Jiang2007}
X.~Jiang, Q.~Xiong, S.~Nam, F.~Qian, Y.~Li, C.~M. Lieber, {InAs/InP Radial
  Nanowire Heterostructures as High Electron Mobility Devices}, Nano Letters
  7~(10) (2007) 3214--3218.
\newblock \href {http://dx.doi.org/10.1021/nl072024a}
  {\path{doi:10.1021/nl072024a}}.

\bibitem{Ford2009}
A.~C. Ford, J.~C. Ho, Y.-L. Chueh, Y.-C. Tseng, Z.~Fan, J.~Guo, J.~Bokor,
  A.~Javey, {Diameter-Dependent Electron Mobility of InAs Nanowires}, Nano
  Letters 9~(1) (2009) 360--365.
\newblock \href {http://dx.doi.org/10.1021/nl803154m}
  {\path{doi:10.1021/nl803154m}}.

\bibitem{Tilburg2010}
J.~W.~W. van Tilburg, R.~E. Algra, W.~G.~G. Immink, M.~Verheijen, E.~P. A.~M.
  Bakkers, L.~P. Kouwenhoven, {Surface passivated InAs/InP core/shell
  nanowires}, Semiconductor Science and Technology 25 (2010) 024011.
\newblock \href {http://dx.doi.org/10.1088/0268-1242/25/2/024011}
  {\path{doi:10.1088/0268-1242/25/2/024011}}.

\bibitem{Zanolli2007}
Z.~Zanolli, M.-E. Pistol, L.~E. Fröberg, L.~Samuelson, {Quantum-confinement
  effects in InAs–InP core–shell nanowires}, Journal of Physics: Condensed
  Matter 19~(29) (2007) 295219.
\newblock \href {http://dx.doi.org/10.1088/0953-8984/19/29/295219}
  {\path{doi:10.1088/0953-8984/19/29/295219}}.

\bibitem{Borgstrom2005}
M.~T. Borgström, V.~Zwiller, E.~Müller, A.~Imamoglu, {Optically Bright
  Quantum Dots in Single Nanowires}, Nano Letters 5~(7) (2005) 1439--1443.
\newblock \href {http://dx.doi.org/10.1021/nl050802y}
  {\path{doi:10.1021/nl050802y}}.

\bibitem{Mohan2006}
P.~Mohan, J.~Motohisa, T.~Fukui, {Fabrication of InP/InAs/InP core-multishell
  heterostructure nanowires by selective area metalorganic vapor phase
  epitaxy}, Applied Physics Letters 88~(13) (2006) 133105.
\newblock \href {http://dx.doi.org/10.1063/1.2189203}
  {\path{doi:10.1063/1.2189203}}.

\bibitem{Masumoto2011}
Y.~Masumoto, Y.~Hirata, P.~Mohan, J.~Motohisa, T.~Fukui, {Polarized
  photoluminescence from single wurtzite InP/InAs/InP core-multishell
  nanowires}, Applied Physics Letters 98~(21) (2011) 211902.
\newblock \href {http://dx.doi.org/10.1063/1.3592855}
  {\path{doi:10.1063/1.3592855}}.

\bibitem{Masumoto2010}
Y.~Masumoto, K.~Goto, B.~Pal, P.~Mohan, J.~Motohisa, T.~Fukui, {Spectral
  diffusion of type-II excitons in InP/InAs/InP core-multishell nanowires},
  Physica E: Low-dimensional Systems and Nanostructures 42~(10) (2010) 2579 --
  2582, 14th International Conference on Modulated Semiconductor Structures.
\newblock \href {http://dx.doi.org/10.1016/j.physe.2009.10.065}
  {\path{doi:10.1016/j.physe.2009.10.065}}.

\bibitem{Li2007}
H.-Y. Li, O.~Wunnicke, M.~T. Borgström, W.~G.~G. Immink, M.~H.~M. van Weert,
  M.~A. Verheijen, E.~P. A.~M. Bakkers, {Remote p-Doping of InAs Nanowires},
  Nano Letters 7~(5) (2007) 1144--1148.
\newblock \href {http://dx.doi.org/10.1021/nl0627487}
  {\path{doi:10.1021/nl0627487}}.

\bibitem{Tchernycheva2007}
M.~Tchernycheva, G.~E. Cirlin, G.~Patriarche, L.~Travers, V.~Zwiller,
  U.~Perinetti, J.-C. Harmand, {Growth and Characterization of InP Nanowires
  with InAsP Insertions}, Nano Letters 7~(6) (2007) 1500--1504.
\newblock \href {http://dx.doi.org/10.1021/nl070228l}
  {\path{doi:10.1021/nl070228l}}.

\bibitem{Fortuna2010}
S.~A. Fortuna, X.~Li, {Metal-catalyzed semiconductor nanowires: a review on the
  control of growth directions}, Semiconductor Science and Technology 25~(2)
  (2010) 024005.
\newblock \href {http://dx.doi.org/10.1088/0268-1242/25/2/024005}
  {\path{doi:10.1088/0268-1242/25/2/024005}}.

\bibitem{Haapamaki2011}
C.~M. Haapamaki, R.~R. LaPierre, {Mechanisms of molecular beam epitaxy growth
  in InAs/InP nanowire heterostructures}, Nanotechnology 22~(33) (2011) 335602.
\newblock \href {http://dx.doi.org/10.1088/0957-4484/22/33/335602}
  {\path{doi:10.1088/0957-4484/22/33/335602}}.

\bibitem{Plante2009}
M.~C. Plante, R.~R. LaPierre, {Analytical description of the metal-assisted
  growth of III-V nanowires: Axial and radial growths}, Applied Physics Letters
  105~(11) (2009) 114304.
\newblock \href {http://dx.doi.org/10.1063/1.3131676}
  {\path{doi:10.1063/1.3131676}}.

\bibitem{Raychaudhuri2006}
S.~Raychaudhuri, E.~T. Yu, {Critical dimensions in coherently strained coaxial
  nanowire heterostructures}, Journal of Applied Physics 99~(11) (2006) 114308.
\newblock \href {http://dx.doi.org/10.1063/1.2202697}
  {\path{doi:10.1063/1.2202697}}.

\bibitem{Trammell2008}
T.~E. Trammell, X.~Zhang, Y.~Li, L.-Q. Chen, E.~C. Dickey, {Equilibrium
  strain-energy analysis of coherently strained core–shell nanowires},
  Journal of Crystal Growth 310~(12) (2008) 3084 -- 3092.
\newblock \href {http://dx.doi.org/10.1016/j.jcrysgro.2008.02.037}
  {\path{doi:10.1016/j.jcrysgro.2008.02.037}}.

\bibitem{Raychaudhuri2006JVSTB}
S.~Raychaudhuri, E.~T. Yu, {Calculation of critical dimensions for wurtzite and
  cubic zinc blende coaxial nanowire heterostructures}, Vol.~24, AVS, 2006, pp.
  2053--2059.
\newblock \href {http://dx.doi.org/10.1116/1.2216715}
  {\path{doi:10.1116/1.2216715}}.

\bibitem{Kastner2004}
G.~K\"{a}stner, U.~G\"{o}sele, {Stress and dislocations at cross-sectional
  heterojunctions in a cylindrical nanowire}, Philosophical Magazine 84~(35)
  (2004) 3803--3824.
\newblock \href {http://dx.doi.org/10.1080/1478643042000281389}
  {\path{doi:10.1080/1478643042000281389}}.

\bibitem{Gutkin2011}
M.~Y. Gutkin, K.~V. Kuzmin, A.~G. Sheinerman, {Misfit stresses and relaxation
  mechanisms in a nanowire containing a coaxial cylindrical inclusion of finite
  height}, Physica Status Solidi B 248~(7) (2011) 1651--1657.
\newblock \href {http://dx.doi.org/10.1002/pssb.201046452}
  {\path{doi:10.1002/pssb.201046452}}.

\bibitem{Fang2009}
Q.~Fang, H.~Song, Y.~Liu, {Misfit dislocations in an annular film grown on a
  cylindrical nanowire with different elastic constants}, Physica B: Condensed
  Matter 404~(14-15) (2009) 1897 -- 1900.
\newblock \href {http://dx.doi.org/10.1016/j.physb.2008.12.009}
  {\path{doi:10.1016/j.physb.2008.12.009}}.

\bibitem{Vurgaftman2001}
I.~Vurgaftman, J.~R. Meyer, L.~R. Ram-Mohan, {Band parameters for III-V
  compound semiconductors and their alloys}, Journal of Applied Physics 89~(11)
  (2001) 5815--5875.
\newblock \href {http://dx.doi.org/10.1063/1.1368156}
  {\path{doi:10.1063/1.1368156}}.

\bibitem{Paladugu2008}
M.~Paladugu, J.~Zou, Y.-N. Guo, X.~Zhang, H.~J. Joyce, Q.~Gao, H.~H. Tan,
  C.~Jagadish, Y.~Kim, {Polarity driven formation of InAs/GaAs hierarchical
  nanowire heterostructures}, Applied Physics Letters 93~(20) (2008) 201908.
\newblock \href {http://dx.doi.org/10.1063/1.3033551}
  {\path{doi:10.1063/1.3033551}}.

\bibitem{Paladugu2009}
M.~Paladugu, J.~Zou, Y.~Guo, X.~Zhang, H.~Joyce, Q.~Gao, H.~Tan, C.~Jagadish,
  Y.~Kim, {Evolution of Wurtzite Structured GaAs Shells Around InAs Nanowire
  Cores}, Nanoscale Research Letters 4~(8) (2009) 846--849.
\newblock \href {http://dx.doi.org/10.1007/s11671-009-9326-6}
  {\path{doi:10.1007/s11671-009-9326-6}}.

\bibitem{Kavanaugh2011}
K.~L. Kavanagh, J.~Salfi, I.~Savelyev, M.~Blumin, H.~E. Ruda, {Transport and
  strain relaxation in wurtzite InAs?GaAs core-shell heterowires}, Applied
  Physics Letters 98~(15) (2011) 152103.
\newblock \href {http://dx.doi.org/10.1063/1.3579251}
  {\path{doi:10.1063/1.3579251}}.

\bibitem{Keplinger2009}
M.~Keplinger, T.~Mårtensson, J.~Stangl, E.~Wintersberger, B.~Mandl,
  D.~Kriegner, V.~Holý, G.~Bauer, K.~Deppert, L.~Samuelson, {Structural
  Investigations of Core−shell Nanowires Using Grazing Incidence X-ray
  Diffraction}, Nano Letters 9~(5) (2009) 1877--1882.
\newblock \href {http://dx.doi.org/10.1021/nl803881b}
  {\path{doi:10.1021/nl803881b}}.

\bibitem{Dick2010}
K.~A. Dick, P.~Caroff, J.~Bolinsson, M.~E. Messing, J.~Johansson, K.~Deppert,
  L.~R. Wallenberg, L.~Samuelson, {Control of III-V nanowire crystal structure
  by growth parameter tuning}, Semiconductor Science and Technology 25~(2)
  (2010) 024009.
\newblock \href {http://dx.doi.org/10.1088/0268-1242/25/2/024009}
  {\path{doi:10.1088/0268-1242/25/2/024009}}.

\bibitem{Plante2008}
M.~Plante, R.~LaPierre, {Au-assisted growth of GaAs nanowires by gas source
  molecular beam epitaxy: Tapering, sidewall faceting and crystal structure},
  Journal of Crystal Growth 310~(2) (2008) 356 -- 363.
\newblock \href {http://dx.doi.org/10.1016/j.jcrysgro.2007.10.050}
  {\path{doi:10.1016/j.jcrysgro.2007.10.050}}.

\bibitem{Plante2008Nt}
M.~C. Plante, R.~R. LaPierre, {Control of GaAs nanowire morphology and crystal
  structure}, Nanotechnology 19~(49) (2008) 495603.
\newblock \href {http://dx.doi.org/10.1088/0957-4484/19/49/495603}
  {\path{doi:10.1088/0957-4484/19/49/495603}}.

\bibitem{Fakhr2010}
A.~Fakhr, Y.~M. Haddara, R.~R. LaPierre, {Dependence of InGaP nanowire
  morphology and structure on molecular beam epitaxy growth conditions},
  Nanotechnology 21~(16) (2010) 165601.
\newblock \href {http://dx.doi.org/10.1088/0957-4484/21/16/165601}
  {\path{doi:10.1088/0957-4484/21/16/165601}}.

\bibitem{Chen2007}
C.~Chen, S.~Shehata, C.~Fradin, R.~LaPierre, C.~Couteau, G.~Weihs,
  {Self-Directed Growth of AlGaAs Core−Shell Nanowires for Visible Light
  Applications}, Nano Letters 7~(9) (2007) 2584--2589.
\newblock \href {http://dx.doi.org/10.1021/nl070874k}
  {\path{doi:10.1021/nl070874k}}.

\bibitem{Chen2008}
C.~Chen, N.~Braidy, C.~Couteau, C.~Fradin, G.~Weihs, R.~LaPierre, {Multiple
  Quantum Well AlGaAs Nanowires}, Nano Letters 8~(2) (2008) 495--499.
\newblock \href {http://dx.doi.org/10.1021/nl0726306}
  {\path{doi:10.1021/nl0726306}}.

\bibitem{OkamotoAuIn}
H.~Okamoto, Au-in (gold-indium), {Journal of Phase Equilibria and Diffusion} 25
  (2004) 197--198.
\newblock \href {http://dx.doi.org/10.1007/s11669-004-0029-5}
  {\path{doi:10.1007/s11669-004-0029-5}}.

\bibitem{OkamotoAlAu}
H.~Okamoto, Al-au (aluminum-gold), {Journal of Phase Equilibria and Diffusion}
  26 (2005) 391--393.
\newblock \href {http://dx.doi.org/10.1007/s11669-005-0098-0}
  {\path{doi:10.1007/s11669-005-0098-0}}.

\bibitem{Grandjean1996}
N.~Grandjean, J.~Massies, {Kinetics of surfactant-mediated epitaxy of III-V
  semiconductors}, Phys. Rev. B 53 (1996) R13231--R13234.
\newblock \href {http://dx.doi.org/10.1103/PhysRevB.53.R13231}
  {\path{doi:10.1103/PhysRevB.53.R13231}}.

\bibitem{Tournie1995}
E.~Tourni\'{e}, N.~Grandjean, A.~Trampert, J.~Massies, K.~Ploog,
  {Surfactant-mediated molecular-beam epitaxy of III–V strained-layer
  heterostructures}, Journal of Crystal Growth 150, Part 1~(0) (1995) 460 --
  466.
\newblock \href {http://dx.doi.org/10.1016/0022-0248(95)80254-A}
  {\path{doi:10.1016/0022-0248(95)80254-A}}.

\bibitem{Vasallo2007}
B.~Vasallo, N.~Wichmann, S.~Bollaert, Y.~Roelens, A.~Cappy, T.~Gonzalez,
  D.~Pardo, J.~Mateos, {Comparison Between the Dynamic Performance of Double-
  and Single-Gate AlInAs/InGaAs HEMTs}, Electron Devices, IEEE Transactions on
  54~(11) (2007) 2815 --2822.
\newblock \href {http://dx.doi.org/10.1109/TED.2007.907801}
  {\path{doi:10.1109/TED.2007.907801}}.

\bibitem{Pistol2008}
M.-E. Pistol, C.~E. Pryor, {Band structure of core-shell semiconductor
  nanowires}, Phys. Rev. B 78 (2008) 115319.
\newblock \href {http://dx.doi.org/10.1103/PhysRevB.78.115319}
  {\path{doi:10.1103/PhysRevB.78.115319}}.

\end{thebibliography}
\end{document}